# The formation of Jupiter by hybrid pebble-planetesimal accretion

**Author:** Yann Alibert[1], Julia Venturini[2], Ravit Helled[2], Sareh Ataiee[1], Remo Burn[1], Luc Senecal[1], Willy Benz[1], Lucio Mayer[2], Christoph Mordasini[1], Sascha P. Quanz[3], Maria Schönbächler[4]

**Affiliations:**

[1]Physikalisches Institut & Center for Space and Habitability, Universität Bern, Gesellschaftsstrasse 6, 3012 Bern, Switzerland

[2]Institut for Computational Sciences, Universität Zürich, Winterthurstrasse 190, 8057 Zürich, Switzerland

[3] Institute for Particle Physics and Astrophysics, ETH Zürich, Wolfgang-Pauli-Strasse 27, 8093 Zürich, Switzerland

[4]Institute of Geochemistry and Petrology, ETH Zürich, Clausiusstrasse 25, 8092 Zürich, Switzerland

**The standard model for giant planet formation is based on the accretion of solids by a growing planetary embryo, followed by rapid gas accretion once the planet exceeds a so-called critical mass[1]. The dominant size of the accreted solids (cm-size particles named *pebbles* or km to hundred km-size bodies named *planetesimals*) is, however, unknown[1,2]. Recently, high-precision measurements of isotopes in meteorites provided evidence for the existence of two reservoirs in the early Solar System[3]. These reservoirs remained separated from ~1 until ~ 3 Myr after the beginning of the Solar System's formation. This separation is interpreted as resulting from Jupiter growing and becoming a barrier for material transport. In this framework, Jupiter reached ~20 Earth masses ($M_\oplus$) within ~1 Myr and slowly grew to ~50 $M_\oplus$ in the subsequent 2 Myr before reaching its present-day mass[3]. The evidence that Jupiter slowed down its growth after reaching 20 $M_\oplus$ for at least 2 Myr is puzzling because a planet of this mass is expected to trigger fast runaway gas accretion[4,5]. Here, we use theoretical models to describe the conditions allowing for such a slow accretion and show that Jupiter grew in three distinct phases. First, rapid pebble accretion brought the major part of Jupiter's core mass. Second, slow planetesimal accretion provided the energy required to hinder runaway gas accretion during 2 Myr. Third, runaway gas accretion proceeded. Both pebbles and planetesimals therefore have an important role in Jupiter's formation.**

High-precision measurements of isotopes (Mo, W and Pt) in meteorites have recently been used to temporally and spatially constrain the early Solar System, by combining two main cosmochemical observations (see Methods). First, cosmochemical data of the youngest inclusions (named chondrules) in primitive meteorites constrain their accretion age. Second, distinct nucleosynthetic isotope compositions (e.g. in Mo or W) that were imprinted in dust accreted by growing bodies allow to identify regions in the protoplanetary disk with distinct dust compositions. Based on these data, the existence of two main reservoirs of small bodies that existed in the early Solar System can be infered[6,7,8]. These reservoirs remained well-separated for a period of about ~2 Myr, likely because of the formation of Jupiter[3]. These cosmochemical evidence, which were never included in Jupiter's growth models, place severe constraints on planetary formation models.

We simulate Jupiter's growth by solid and gas accretion using state-of-the art planet formation models[9] to determine the time required for Jupiter to reach 50 $M_\oplus$, assuming Jupiter formed in situ. We consider different values for the mass of Jupiter at 1 Myr, and for the average accretion rate of solids after 1 Myr. As the opacity and the composition of Jupiter's envelope are not precisely known, we ran models using a large range of assumptions (low or large opacity, pure Hydrogen-Helium or envelope enriched in heavier elements). Model results show that the cosmochemical constraints are met, but only with a planet mass at 1 Myr between ~5 and 16 $M_\oplus$ depending on the assumed conditions (Fig. 1). Therefore, the minimum mass of the forming Jupiter that is required to prevent the transport of pebbles (the so-called *pebble isolation mass*), is somehow smaller than the 20 $M_\oplus$ quoted above. Note that the precise value of the pebble isolation mass and the mass that Jupiter should have attained at time ~3 Myr after the beginning of the Solar System are not directly derived from cosmochemical studies, but result from theoretical interpretation[3].

Our models also show that a relatively high solid accretion rate (at least $10^{-6}$ $M_\oplus$/yr) is required to prevent rapid gas accretion after 1 Myr. Indeed, slow gas accretion is only possible through a significant thermal support of the gas-dominated envelope that can counteract the strong gravity of the planetary core. We find that the dissipation of the kinetic energy from in-falling solids thermally supports the envelope and inhibits high gas accretion rates. We checked that the ranges of values of pebble isolation mass and solid accretion rates are very robust and insensitive to the envelope composition and/or the opacity values, planet's location, and disk properties (see Supplementary figures 1 and 2 in Supplementary Information).

Since Jupiter reached the pebble isolation mass around ~1 Myr, maintaining a high solid accretion rate beyond this time must result from the accretion of planetesimals, which do not feel the isolating effect of the planet as pebble do (see Supplementary Information). During the first Myr the solid accretion rate needs to be as high as ~ $10^{-5}$ $M_\oplus$/yr in order for Jupiter to reach a mass of ~5 to 16 $M_\oplus$ in only one Myr. This accretion rate is too high to result from the accretion of planetesimals, and must result from the accretion of pebbles (see Supplementary Information). On the other hand, a rate of at least $10^{-6}$ $M_\oplus$/ yr in planetesimal accretion is required to stall gas runaway accretion and keep the planetary mass below 50 $M_\oplus$ for the next 2 Myr. Hence, fulfilling the cosmochemical time constraints[3] in a Jupiter formation scenario is only possible through a hybrid accretion process, where, first, pebbles provide high accretion rates and grow a large core (~5 to 16 $M_\oplus$), and second, significant planetesimal accretion sets in afterwards. This planetesimal accretion, which occurs after 1 Myr, supplies the energy required for delaying rapid gas accretion, and only modestly contributes to the core's mass.

The derived accretion rate of planetesimals onto Jupiter represents a significant flux of infalling solids. Such high accretion rates cannot be sustained by large (hundreds of km in size) planetesimals given the excitation they experience from the gravitational interaction with a growing planetary embryo and the inability of gas drag to damp the eccentricity and inclination of such big objects[10,11]. Thus, our results suggest that a significant mass of small planetesimals (km in size) was present in the Solar Nebula at 1 Myr (see Fig. 2 and Supplementary Information) in apparent contradiction to recent studies suggesting the existence of large primordial planetesimals[12,13]. These smaller objects would, therefore, be second-generation planetesimals, resulting from the fragmentation of larger primordial objects[14]. Indeed, the presence of a planet of a few $M_\oplus$ leads to collisions that are violent enough to disrupt primordial planetesimals[14]. Moreover, the collision timescale among large planetesimals is short enough to allow the formation of small ones by fragmentation in less than 1 Myr (see Supplementary Information). In this way, the initial growth of Jupiter by pebble accretion during the first Myr provided the conditions to fragment large primordial planetesimals into small second-generation objects in a timely manner.

Our formation scenario also provides a solution to the problem of timing of pebble accretion. Indeed, pebble accretion is so efficient that objects become quickly more massive than Jupiter unless accretion starts shortly before the dispersal of the protoplanetary disk[15,16]. This timing is inconsistent with detailed models of pebble growth, which conclude that pebbles form and accrete early[17]. In the hybrid pebble-planetesimal scenario, the formation of Jupiter-mass planets is stretched over a few Myrs, comparable to the typical lifetimes of circumstellar disks[18]. In this case, it is possible that pebbles are accreted in the early phases of the protoplanetary disk evolution, without leading necessarily to the formation of massive planets.

We conclude that Jupiter formed in a 3-step process (Fig. 3): (1) Jupiter's core grew by pebble accretion. The contribution of primordial large planetesimals to the solid accretion was negligible. As Jupiter's core became more massive, large primordial planetesimals dynamically heated up, collided and formed second generation smaller sized planetesimals. (2) Pebble accretion ceased (Jupiter reached the pebble isolation mass) and the protoplanet grew more slowly by accretion of small planetesimals. The solid accretion rate remained high enough to provide sufficient thermal support to the gas envelope and to prevent rapid gas accretion. (3) The critical mass for gas accretion was reached, gas rapidly accreted and Jupiter reached its present-day mass. During this last phase, further solids may have been accreted increasing the final heavy-element content in Jupiter[19].

Our simulations show that the total heavy-element mass in Jupiter (core and envelope) prior to runaway gas accretion (accounting for both pebble and planetesimal accretion) ranges from 6 to 20 $M_\oplus$. This value can be compared with the Jupiter's heavy-element mass as derived from structure models, which ranges from 23.6 to 46.2 $M_\oplus$[20]. This comparison implies that Jupiter accreted up to ~25 $M_\oplus$ during runaway gas accretion or at a later stage[19]. Heavy elements that accreted late do not necessarily reach the core. They can dissolve in the envelope[21], leading to envelope enrichment and the formation of heavy element gradients[22].

In this new hybrid pebble-planetesimal scenario, the time a protoplanet spends in the mass range of 15-50 $M_\oplus$ extends over a few Myr before rapid gas accretion takes place. Since the final mass of a planet is determined by the dissipation of the protoplanetary disk, our new formation scenario increases the likelihood of forming intermediate-mass planets, and this provides a natural explanation for the formation of Uranus and Neptune[1,23].

This work has been developed in the framework of the National Center for Competence in Research PlanetS funded by the Swiss National Science Foundation (SNSF). YA, WB, RB acknowledge support from the Swiss National Science Foundation (SNSF) under grant 200020_172746. CM acknowledges the support from the Swiss National Science Foundation (SNSF) under grant BSSGI0_155816 "PlanetsInTime". RH acknowledges support from the Swiss National Science Foundation (SNSF) project 200021_169054. YA acknowledges the support of the European Research Council under grant 239605 "PLANETOGENESIS". MS acknowledges the support of the European Research Council under the European Union's Seventh Framework Programme (FP7/2007–2013)/ERC Grant agreement no. [279779]. We thank Clement Surville for sharing results of hydrodynamical simulations of disk-planet interaction prior to publication.

**Author contributions**: YA initiated the project. YA, JV, SA, RB and LS performed the theoretical calculations. YA, JV and RH led the writing of the manuscript, and all authors contributed to the discussion and the interpretation of the results.

**Correspondence**: Correspondence should be addressed to Yann Alibert (alibert@space.unibe.ch)

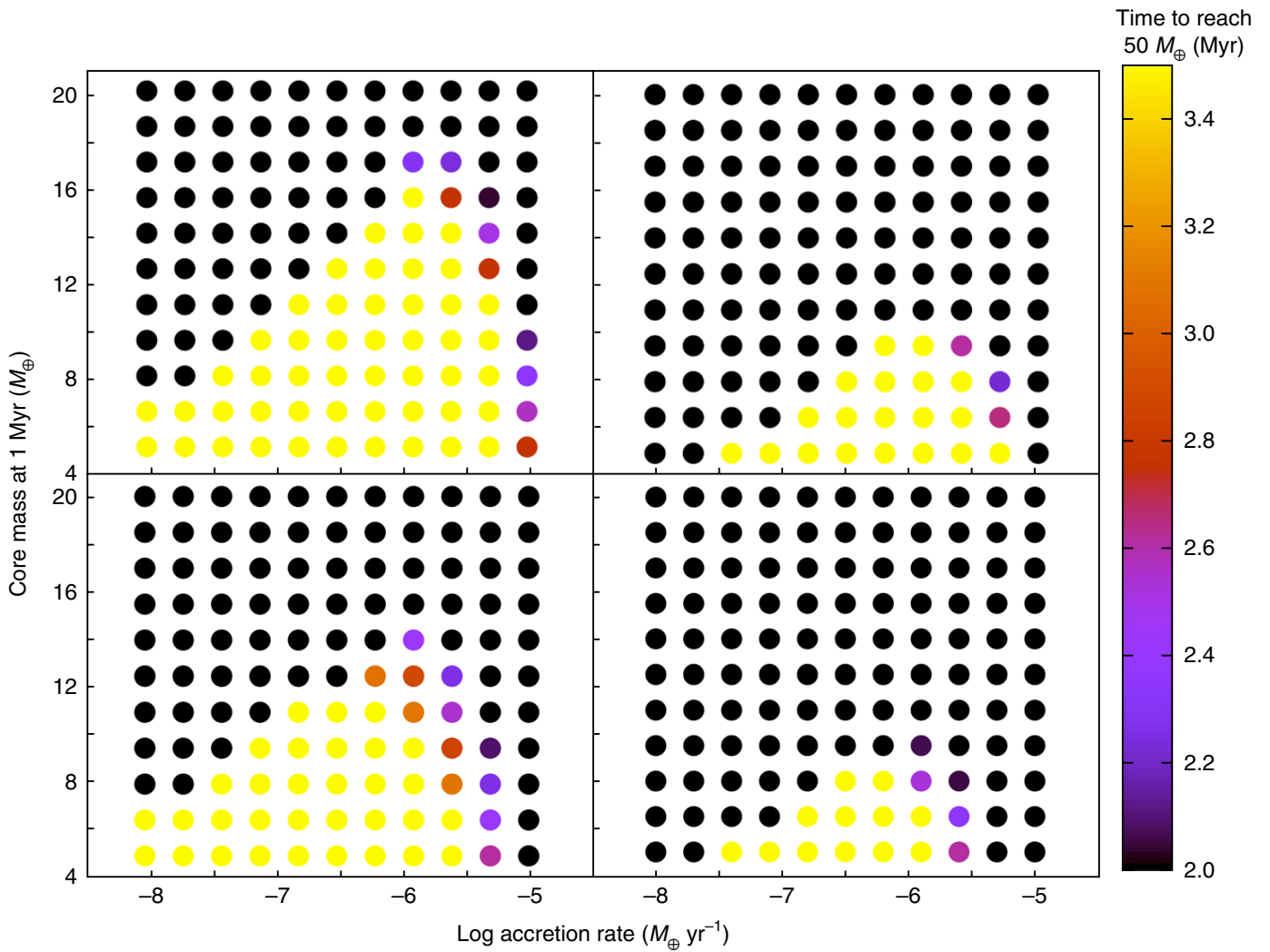

*Fig. 1. Time to reach 50 $M_\oplus$ as a function of the core mass at 1 Myr and the solid accretion rate (in Earth masses per year, log scale).* The yellow region delimits the part of the diagram where the core growth is too slow, whereas the black region delimits the part of the diagram where the runaway gas accretion occurs too early (either because the initial core mass is too large or because the heating by incoming planetesimals is too small). The dots with colors between purple and orange indicate the region that is compatible with the growth timescale of Jupiter as obtained from cosmochemical studies[3]. **Upper left**: non-enriched envelope and ISM opacity[24]. **Upper right**: non-enriched envelope and opacity reduced by a factor 10 compared to ISM one. **Lower left**: enriched envelope and ISM opacity. **Lower right**: enriched envelope and reduced opacity. Note that in all cases the parameter space that is consistent with the cosmochemical constraints[3] is rather small.

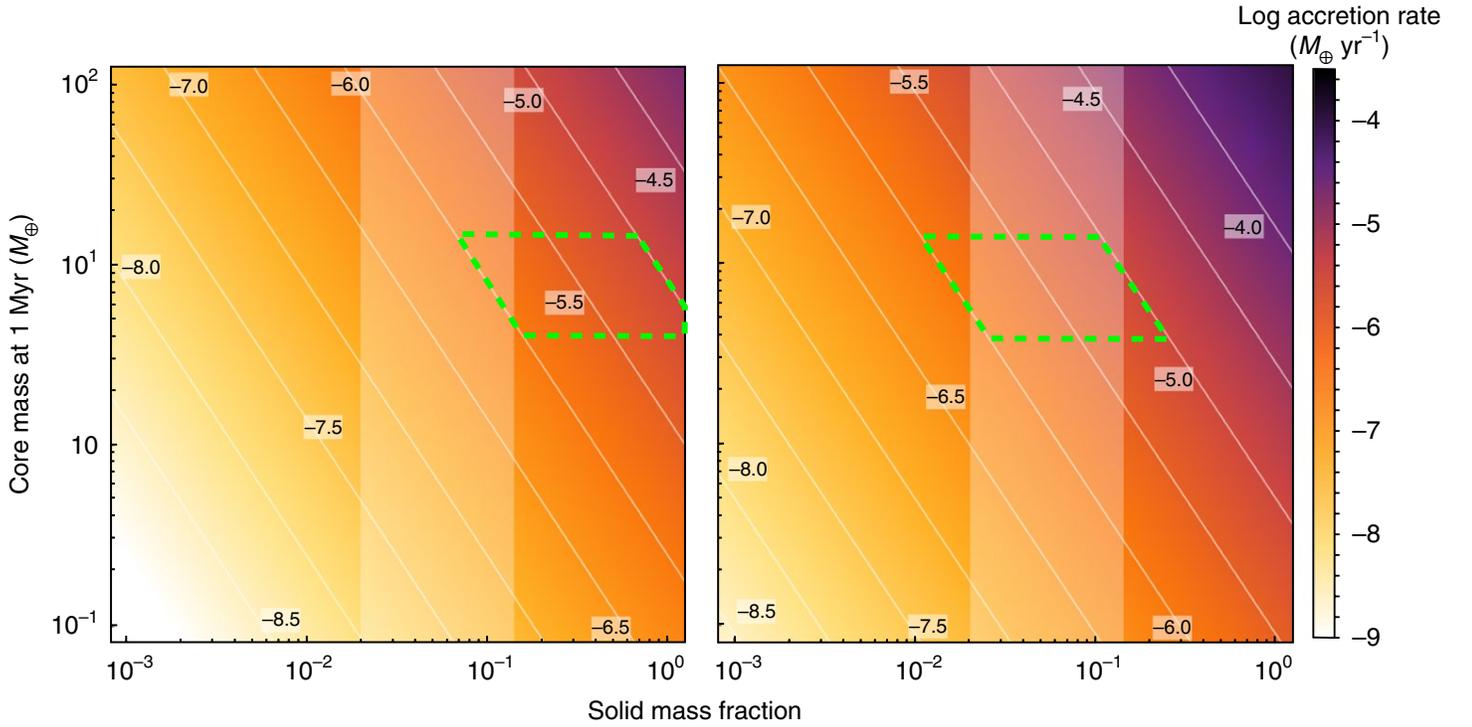

***Fig. 2. Accretion rate of planetesimals as a function of the solid mass fraction and the core mass at 1 Myr.*** *The green region delimits the parameters that allow matching the cosmochemical constraints[3], i.e., a core mass between 4 and ~16 $M_\oplus$ and accretion rate of solids between $10^{-6}$ $M_\oplus$/yr and $10^{-5}$ $M_\oplus$/y). The white shaded area delimits the likely solid mass fraction according to the standard MMSN model for the lowest value[25] and the dust-to-planetesimal formation models for the highest value[26]. The left panel is for large planetesimals (100 km in size), whereas the right panel is for small planetesimals (1 km in size).*

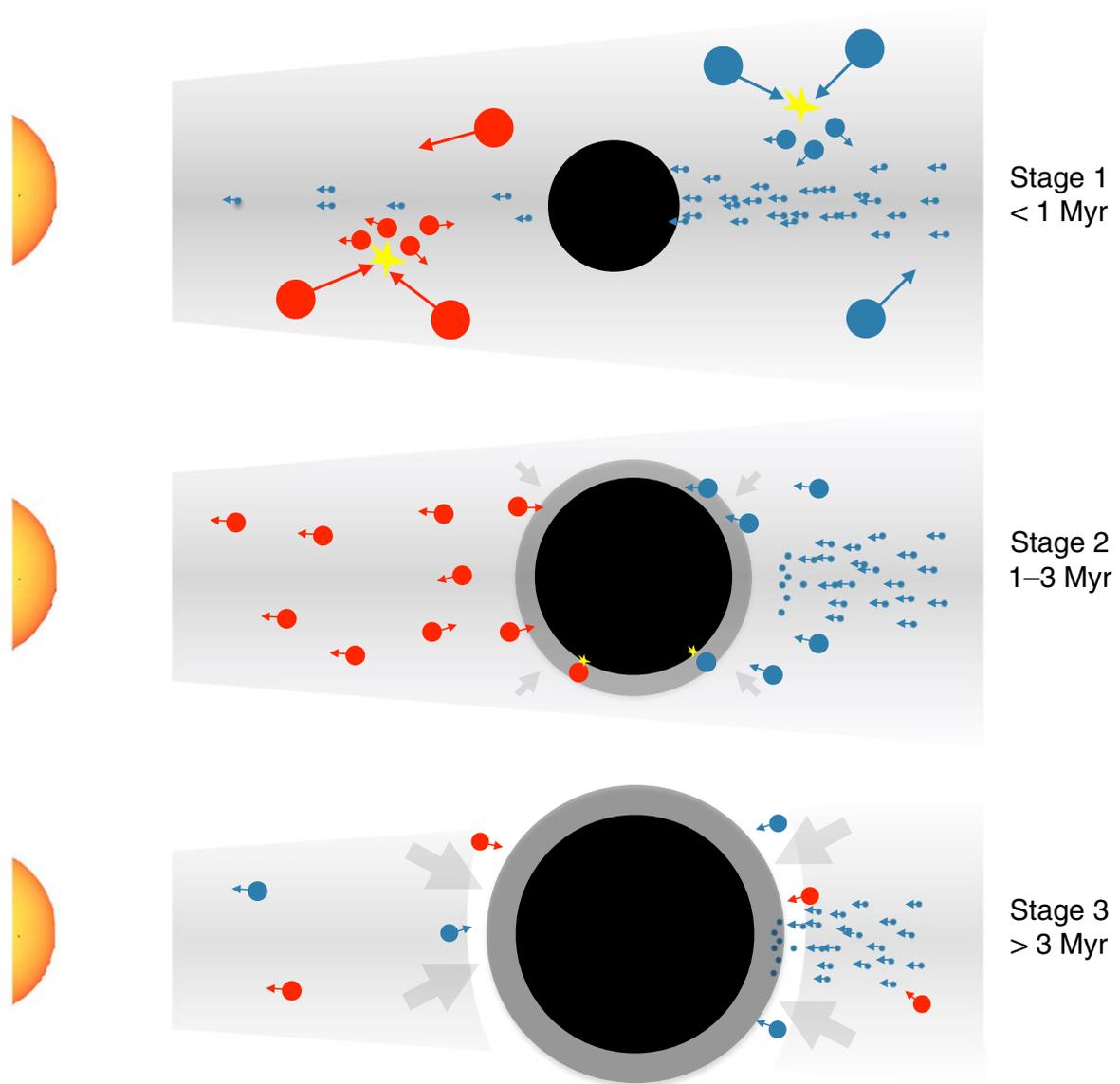

*Fig. 3. The three stages of the hybrid pebble-planetesimal formation model. In **stage 1** (up to 1 Myr), Jupiter grows by pebble accretion (small dots) and planetsimal accretion is negligible. Large primordial planetesimals (large circles) are excited by the growing planet and suffer high collision velocities (large arrows) leading to destructive collisions (yellow) that produce small, second generation planetesimals (small circle). In **stage 2** (1-3 Myr), Jupiter is massive enough to prevent pebble accretion. The energy associated with accretion of small planetesimals is large enough to prevent rapid gas accretion (gray arrows). In **stage 3** (beyond 3 Myr), Jupiter is massive enough to accrete large amounts of gas (H-He). Nearby pebbles and small planetesimals can be gravitationally captured. Ultimately a gap is opened in the Solar nebula stopping further gas accretion. The red and blue colors indicate the two reservoirs (inside Jupiter's orbit and outside Jupiter's orbit respectively) that are separated by Jupiter's growth during stage 2, and get reconnected in stage 3.*

# Methods

## Meteoritic constraints

High-precision measurements of isotopes (Mo, W and Pt) in meteorites have been used by other authors to constrain Jupiter's growth history[3] by combining two main cosmochemical observations. First, cosmochemical data of the youngest inclusions (i.e. chondrules) in primitive meteorites constrain the maximum accretion age for small primitive bodies, while the short-lived $^{182}$Hf-$^{182}$W decay system dates metal-silicate separation and as such the accretion timescales of small differentiated bodies and planets. Second, distinct nucleosynthetic isotope compositions (e.g. in Mo or W) that were imprinted in dust accreted by planetary bodies allow to identify regions in the protoplanetary disk with originally distinct dust compositions. Based on this, cosmochemical data constrain two main reservoirs of small bodies that existed in the early Solar System[4,5,6]. They remained well-separated for a period of about 2-4 Myr[3,27]. The separation of these two reservoirs occurred within the first Myr after the beginning of the Solar System as defined by the formation of the oldest Solar System materials (Calcium Aluminium rich Inclusions (CAIs)). It was proposed[3] that this separation was initiated by the growth of proto-Jupiter reaching pebble isolation mass (20 $M_\oplus$), thereby isolating the population of pebbles inside and outside of its orbit. The two reservoirs remained separated until Jupiter grew massive enough to scatter small bodies, reconnecting the reservoirs. This occurred when Jupiter reached 50 $M_\oplus$, and not earlier than 3-4 Myr after CAI formation[3]. While cosmochemical evidence constrain the timescale of the separation of the reservoirs, it does not constrain the mass that Jupiter had at these epochs.

## Modelling the planetary growth

We compute the planetary growth in the framework of the core accretion model, by solving the planetary internal structure equations[4,5,6], assuming the luminosity results from the accretion of solids and gas contraction.

We consider two limiting cases regarding the fate of solids accreted by proto-Jupiter. In the first case, the so-called non-enriched case, all the accreted heavy elements are assumed to sink to the center (core). In this case the envelope is made of pure H and He. In the second case, the enriched case, we assume that the volatile fraction of the accreted solids is deposited in the envelope, whereas the refractory component reaches the core[20,28]. The volatile fraction is assumed to be 50 wt%, following recent condensation models[29]. The luminosity in this case is the one provided just by the refractory material, since the volatiles are assumed to remain mixed in the envelope and contribute to the luminosity generated by its contraction[20,28]. In all the models presented here, we treat the accretion rate of solids between 1 Myr and 3 Myr as a free parameter that varies from $10^{-8}$ $M_\oplus$/yr to $10^{-5}$ $M_\oplus$/yr.

The internal structure equations[4,5,6] are solved, using as boundary conditions the pressure and temperature in the protoplanetary disk at the position of the planetary embryo and defining the planetary radius as a combination of the Hill and Bondi radii[30]. The evolution of the planetary envelope depends on the used EOS and opacity. For the non-enriched case, we use the EOS of H and He[31]. For the enriched case, the envelope is assumed to be composed of H, He and water, and we take into account the mixture of the three components[28,31,32]. For the opacity, we use either interstellar medium (ISM) opacity[21], or a reduced opacity in which we multiply the ISM opacity by 1/10, in order to mimic the possible opacity reduction due to grain growth[33,34]. The calculations do not include the effect of destruction and replenishment of pebbles in Jupiter's envelope[35,36], since the growth of Jupiter after 1 Myr is dominated by accretion of planetesimals for which the effect of destruction in the planetary envelope is less important[36].

## Disk structure

The disk model provides the pressure and temperature at the formation location of Jupiter, which serve as boundary conditions for the computation of the internal structure. This model is designed to fit 2-D radiative hydrodynamic simulations of protoplanetary disks[37].

## Planetesimal Accretion

In early planet formation models, it was assumed that the accreted solids were large planetesimals[4] (with sizes of hundreds of km), in agreement with several theoretical and observational constraints[9,10]. These planetesimal-based formation models still face the problem that the time required to reach rapid gas accretion is comparable to or even longer than the disk's lifetime[4,15]. This challenge is even more severe if dynamical heating (increased eccentricity and inclination) of the planetesimals by the gravity of a proto-Jupiter is considered[7,38] (see also Supplementary Information) since this hinders the core growth. Dynamical heating is counteracted by damping caused from gas drag and thus primarily affects small planetesimals. Hence, accreting solids of only a few km in size can relieve the timescale problem[7,8]. Numerical simulations, however, predict much larger typical sizes for primordial planetesimal, on the order of tens to hundreds of km[9], with most of the mass stored in the largest bodies, in agreement with the constraints from the asteroid belt[10]. Therefore, km-sized planetesimals are likely generated by collisional fragmentation of large primordial planetesimals. This in turn requires high collision velocities, which results from the gravitational stirring of primordial planetesimals by objects with masses of a few $M_\oplus$[11].

Planetesimal accretion depends on three factors: the amount of planetesimals near the planet, the mass of the forming planet, and the degree of planetesimals excitation. In particular, the planetesimal accretion rate depends on the gravitational focussing factor $F_{grav}$, itself depending inversely on the relative velocity $v_{rel}$ between planetesimals and the growing Jupiter. When planetesimals are dynamically excited (i.e, have large eccentricity and inclination), $v_{rel}$ increases and the planetesimals are accreted less efficiently. Planetesimals are excited by the forming planet and by planetesimal-planetesimal interactions, and also damped by gas drag.

Large planetesimals are more excited than small ones, because gas drag is less active on the formers. Therefore, the relative velocity between planetesimals and the growing Jupiter is larger for large planetesimals, leading to smaller accretion rates than for small planetesimals. We compute the accretion rate of planetesimals[11], for planetesimals of 100 km or 1 km in size, as a function of the planet mass, and planetesimal-to-gas mass ratio. The properties of the gas disk that are required for this calculation (e.g. gas density) are taken form the disk model at a radial distance of 5.2 AU and an age of 1 Myr (when planetesimal accretion begins).

## Fragmentation of large planetesimals

Two conditions are required in order to account for the formation of small planetesimals from the fragmentation of large ones before 1 Myr (when the accretion of pebbles stops). Collisions must be frequent enough (so that small planetesimals are produced rapidly enough), and violent enough (so that collisions lead to fragmentation). We estimate the collision timescale[39,40] between planetesimals of 100 km in size, as a function of the protoplanet's mass, and the solid surface density at 5 AU. The collision frequencies are calculated for a single sized population of planetesimals. The calculation includes the stirring of planetesimals by the growing Jupiter, but not the interaction between planetesimals, which is negligible for planets a few $M_\oplus$ in mass[11]. Including this effect would increase the excitation of planetesimals, leading to even more violent collisions, and further fragmentation. We also include the gas drag that decreases the eccentricity and inclination of planetesimals, and therefore their collision velocity. In order to determine in which case the collisions lead to the destruction of planetesimals, we compared the specific energy of the collision with the one required for disruption $Q^*_D$. We chose for this value a very conservative estimate of $6 \cdot 10^9$ erg/g, which corresponds to the highest value found for any set of compositional parameters[41]. As a result, for all collisions involving an energy larger than $Q^*_D$, planetesimals are expected to be destroyed and to fragment into much smaller objects. More details on the calculation of the fragmentation of large planetesimals is given in the Supplementary Information.

**Data availability statement:** The data that support the plots within this paper and other findings of this study are available from the corresponding author upon reasonable request.

**Additional reference:**